\documentclass[lettersize,journal]{IEEEtran}
    \usepackage{booktabs}
    \usepackage{diagbox}
    \usepackage{makecell}
    \usepackage{array}
    \usepackage{graphicx,amssymb,amsmath}
    \usepackage{multicol}
    \usepackage[noadjust]{cite} 
    \usepackage{setspace}
    \usepackage{subfigure}
    \usepackage{graphicx}
    \usepackage{float}
    \usepackage {url}
    \usepackage{stfloats}
    \usepackage{amsthm,pifont}
    \usepackage{flushend}
    \usepackage{cases,subeqnarray}
    \usepackage{bm,multirow,bigstrut}
    \usepackage{amsmath, amsthm, amssymb}
    \usepackage{textcomp}
    \usepackage{latexsym,bm}
    \usepackage{booktabs}
    \usepackage{mathtools}
    \usepackage{dsfont}
    \usepackage{extarrows}
    \usepackage{epsfig}
    \usepackage{epsfig}
    \usepackage{epstopdf}
    \usepackage[table]{xcolor}
    \usepackage{colortbl} 
    \usepackage{multirow, hhline}
    \usepackage{xcolor}
    \usepackage{array} 
    \usepackage{algorithmicx,algorithm}
    \usepackage{hyperref}

    \theoremstyle{plain}

    \theoremstyle{plain}

    \usepackage{amsmath}

    \IEEEoverridecommandlockouts
\begin{document}
\title{Wireless Agentic AI with Retrieval-Augmented Multimodal Semantic Perception}

\author{Guangyuan Liu, Yinqiu Liu, Ruichen Zhang, Hongyang Du$^{*}$, Dusit Niyato,~\IEEEmembership{Fellow,~IEEE}, Zehui~Xiong,\\ Sumei Sun,~\IEEEmembership{Fellow,~IEEE}, and Abbas Jamalipour,~\IEEEmembership{Fellow,~IEEE},
\thanks{G.~Liu is with the College of Computing and Data Science, the Energy Research Institute @ NTU, Interdisciplinary Graduate Program, Nanyang Technological University, Singapore (e-mail: liug0022@e.ntu.edu.sg).}
\thanks{Y.~Liu, R.~Zhang, D. Niyato are with the College of Computing and Data Science, Nanyang Technological University, Singapore (e-mail: yinqiu001@e.ntu.edu.sg, ruichen.zhang@ntu.edu.sg, dniyato@ntu.edu.sg).}
 \thanks{H.~Du is with the Department of Electrical and Electronic Engineering, University of Hong Kong, Hong Kong (e-mail: duhy@eee.hku.hk).}
\thanks{Z. Xiong is with the Pillar of Information Systems Technology and Design, Singapore University of Technology and Design, Singapore (e-mail: zehui\_xiong@sutd.edu.sg).}
\thanks{S. Sun is with the Institute for Infocomm Research, Agency for Science, Technology and Research, Singapore (e-mail: sunsm@i2r.a-star.edu.sg).}
\thanks{A. Jamalipour is with the School of Electrical and Information Engineering, University of Sydney, Australia (e-mail: a.jamalipour@ieee.org).}
\thanks{* means the corresponding author}%
}
\maketitle
\vspace{-1cm}
\begin{abstract}
The rapid development of multimodal AI and Large Language Models (LLMs) has greatly enhanced real-time interaction, decision-making, and collaborative tasks. However, in wireless multi-agent scenarios, limited bandwidth poses significant challenges to exchanging semantically rich multimodal information efficiently. Traditional semantic communication methods, though effective, struggle with redundancy and loss of crucial details. To overcome these challenges, we propose a Retrieval-Augmented Multimodal Semantic Communication (RAMSemCom) framework. RAMSemCom incorporates iterative, retrieval-driven semantic refinement tailored for distributed multi-agent environments, enabling efficient exchange of critical multimodal elements through local caching and selective transmission. Our approach dynamically optimizes retrieval using deep reinforcement learning (DRL) to balance semantic fidelity with bandwidth constraints. A comprehensive case study on multi-agent autonomous driving demonstrates that our DRL-based retrieval strategy significantly improves task completion efficiency and reduces communication overhead compared to baseline methods.
\end{abstract}


\section{Introduction}

In recent years, Large Language Models (LLMs) have made remarkable strides in natural language comprehension and generation, leading to transformative impacts across diverse fields, including real-time interaction, recommendation systems, and autonomous decision-making~\cite{yu2025benchmarking}. Concurrently, advances in multimodal deep learning have further enhanced the capabilities of AI systems, enabling them to process and integrate multiple modalities such as text, images, and audio. This multimodal intelligence has significantly improved service quality, user experience, and task efficiency in diverse scenarios including autonomous driving and the metaverse.

Building upon these technological developments, the concept of agentic AI has emerged, where intelligent agents operate autonomously and often collaborate to achieve shared or individual objectives~\cite{wei2025internet}. For instance, self-driving cars can coordinate with each other to prevent collisions or to optimize traffic flow, while warehouse robots may jointly manage inventory. Moreover, recent studies discovered that knowledge sharing between LLM agents can complement each other’s strengths, reduce redundant development efforts, and lower overall operational costs~\cite{wei2025internet}. Specifically, the sharing of knowledge among agents can:
\begin{itemize}
    \item \textbf{Enhance Performance and Versatility:} 
    By leveraging diverse models or skill sets, agents collaboratively tackle complex problems that exceed the capabilities of any single model.

    \item \textbf{Improve Resource Efficiency:} 
    Exchanging relevant insights or partial results minimizes duplicate computations, cutting both operational expenses and carbon footprints.

    \item \textbf{Boost Scalability and Accessibility:} 
    Pooling computational resources and expertise allows smaller research teams and developers to engage in advanced AI tasks, fostering a more inclusive ecosystem.
\end{itemize}

In such multi-agent scenarios, bandwidth-limited wireless connections are typically used for exchanging information among agents. These real-time tasks rely on the rapid transmission of critical information over bandwidth-limited wireless channels. Directly exchanging raw sensor data often overwhelms available capacity and includes redundant or low-utility details. Therefore, the communication challenge is not merely to transmit raw bits, but to share semantically rich and concise multimodal information that enables effective coordination. This requirement has propelled research into multimodal semantic communication, an approach dedicated to transmitting only relevant and meaningful content across constrained channels, thereby substantially reducing overhead.

However, existing multimodal semantic communication techniques encounter several critical limitations. Traditional methods~\cite{Semantic}, although more efficient than raw data transmission, often transmit redundant information, leading to inefficient use of bandwidth. Compression-based strategies~\cite{compression} reduce the volume of data but may inadvertently eliminate crucial semantic content, which is particularly detrimental in safety-critical applications such as autonomous driving or real-time monitoring. Moreover, heuristic or rule-based multimodal approaches~\cite{wang2023multimodal} lack adaptability to dynamic contexts and are generally insufficient for handling the complexity inherent in multi-agent collaboration.

In parallel, recent research has demonstrated that LLMs can incorporate external knowledge bases during inference, enabling dynamic retrieval of relevant information as new data emerges~\cite{wang2025retrieval}. This approach eliminates the need to retrain large, monolithic models whenever the underlying knowledge domain evolves. While most existing work has focused on text-based Retrieval-Augmented Generation (RAG), Retrieval-Augmented Perception (RAP) extends these techniques to multimodal contexts \cite{wang2025retrieval}. By selectively integrating critical visual or auditory snippets, RAP can significantly reduce overhead and improve inference efficiency. However, few studies have explored how to adapt RAP-like methods in distributed or multi-agent scenarios.

To address these gaps, this paper proposes a Retrieval-Augmented Multimodal Semantic Communication (RAMSemCom) framework. RAMSemCom facilitates the effective exchange of multimodal knowledge among agentic AI, such as autonomous vehicles, collaborative robots, or edge devices.
Inspired by RAP but tailored for distributed environments, the proposed approach enables each participant to maintain local caches and transmit only the essential portions of visual, textual, or auditory content, thereby enhancing bandwidth efficiency while preserving crucial semantic details.

The main contributions of this paper are as follows:
\begin{itemize}
    \item RAMSemCom framework that systematically incorporates RAP into distributed semantic communication, ensuring minimal but sufficient transmission of critical multimodal elements.
    \item An iterative retrieval and semantic refinement mechanism that dynamically adapts to evolving user demands, preserving essential semantics while avoiding unnecessary overhead.
    \item A comprehensive case study in multi-agent autonomous driving demonstrating the effectiveness of RAMSemCom, highlighting improved task completion efficiency and reduced bandwidth utilization compared to existing baselines.

\end{itemize}

\section{Related Work}

\subsection{Semantic Communication with Multimodal Data}
\begin{figure*}[t]
    \centering
    \includegraphics[width=0.98\linewidth]{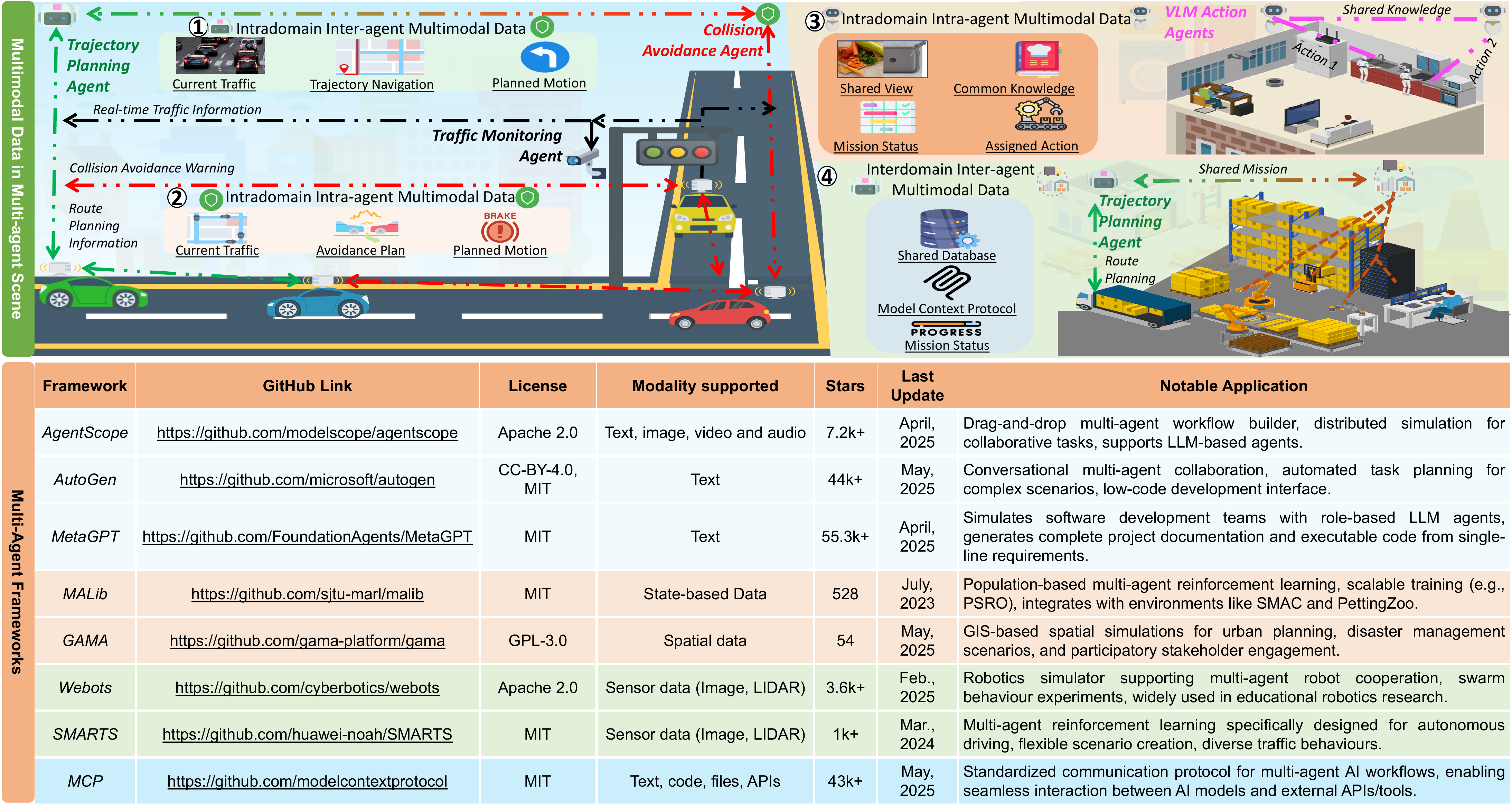}
    \caption{\textbf{Multi-agent Frameworks}: Survey and summary of popular multi-agent frameworks, emphasizing supported modalities, last update status, and notable applications. \textbf{Multimodal Data in Multi-agent Scene}: Illustration of multimodal multi-agent system scenarios demonstrating three common communication patterns: (1) Intradomain inter-agent vehicle coordination, (2) Intradomain intra-agent planning and collaboration (e.g., robot collaboration scenario by Helix AI at https://www.figure.ai/news/helix), and (3) Interdomain inter-agent logistics and transportation coordination.}
    \label{fig:semantic_comm_overview}
\end{figure*}
In modern 6G networks, semantic communication aims to exchange essential meaning from heterogeneous data sources (e.g., text, images, audio) rather than raw bits. Traditional methods focus on deep compression such as end-to-end models that jointly learn source–channel encoders can outperform separate codecs in noisy conditions while greatly reducing bandwidth~\cite{liu2024semantic}. However, these methods often necessitate retraining when encountering new data domains and may inadvertently discard critical fine-grained details. Alternatively, selective transmission strategies, which transmit only salient features such as bounding boxes or key audio cues, further reduce bandwidth consumption by dynamically reconstructing or refining omitted details at the receiver side~\cite{gu2023semantic}. Despite their bandwidth efficiency, these methods depend heavily on robust reconstruction models and risk losing subtle contextual information that might not have been selected during transmission.

Recent multimodal semantic communication studies increasingly leverage advanced multimodal generative models to enhance communication efficiency in bandwidth-constrained environments. For instance, TokCom~\cite{qiao2025token} employs a context-aware approach where a pretrained vision-language model infers missing content from transmitted token representations, thereby reducing the need for retransmissions. Similarly, Guo et al.~\cite{guo2025hierarchy} use a hierarchical attention mechanism to adaptively transmit only the most relevant semantic features based on real-time channel conditions in multimodal data fusion tasks. These methods demonstrate the promise of adaptive context-awareness in reducing transmission overhead while maintaining task fidelity.
However, most existing works remain confined to specific, task-oriented scenarios under controlled conditions such as fixed image-question pairs~\cite{xie2022task} or synthetic dataset evaluations. They often lack mechanisms for dynamically incorporating external knowledge or refining transmissions based on evolving receiver requirements. For example, while TokCom exploits context inference, it cannot retrieve new information beyond the model’s pretrained scope. Likewise, many approaches do not support iterative refinement of multimodal content based on semantic relevance, limiting their applicability in dynamic, real-world or multi-agent environments where data distributions and task demands continually shift. This highlights a critical gap in the current landscape and motivates the need for flexible, retrieval-augmented semantic communication frameworks.

{

\subsection{Retrieval-Augmented Everything}\label{subsec:rag}

RAG has become a prominent method to improve the factual accuracy and adaptability of LLMs. Instead of relying solely on parametric memory, RAG dynamically queries external data sources during inference, grounding model outputs in up-to-date and context-relevant information~\cite{gao2023retrieval}. To clarify its architecture and broader relevance, we briefly unpack the three key components:

\begin{itemize}
    \item \textbf{Retrieval} – Refers to the act of querying an external knowledge base, document store, or vector database using a latent representation of the current input. This retrieval is typically based on semantic similarity, enabling the model to access pertinent facts or context during runtime.

    \item \textbf{Augmented} – Once the information is retrieved, it is appended to or integrated into the model’s input context, effectively enriching the prompt with non-parametric memory. This augmentation enhances reasoning and factual accuracy without modifying model weights.

    \item \textbf{Generation} – Finally, the LLM processes the augmented context to generate a response that is both fluent and grounded in the retrieved content. This step preserves the model’s generative flexibility while improving relevance and traceability.
\end{itemize}

Most RAG implementations have focused on text-based tasks such as open-domain question answering, document summarization, and multi-turn dialogue~\cite{gao2023retrieval}. However, this paradigm has been extended to vision-centric applications. For example, RAP~\cite{wang2025retrieval} applies similar principles to image understanding: instead of processing entire high-resolution inputs, the system selectively retrieves a set of task-relevant image patches based on semantic similarity. This strategy reduces computation and bandwidth while preserving essential visual details, making RAP particularly useful in constrained environments.

In wireless edge AI and Internet-of-Things (IoT) scenarios, retrieval-enhanced methods allow intelligent agents (e.g, drones, autonomous vehicles, or distributed sensors) to formulate semantic summaries or fetch contextual data before initiating transmission. In multi-agent systems, each agent can access shared repositories to obtain missing task-related information (e.g., prior observations, maps, or annotated examples) without exchanging full-resolution sensor data. This selective and context-aware retrieval mechanism significantly reduces communication overhead while preserving the semantic richness necessary for coordinated and intelligent decision-making.

The growing popularity of retrieval-augmented architectures thus reflects a broader shift toward dynamic, context-sensitive inference pipelines. These pipelines prioritize the transmission or processing of only the most task-relevant data, setting the stage for scalable, bandwidth-efficient systems in complex and distributed environments.

\subsection{Multi-Agent AI Systems and Knowledge Sharing}

Traditionally, autonomous agents operating in shared environments rely on exchanging raw sensor outputs or detailed feature-level data among peers. However, this approach quickly saturates wireless bandwidth, making it inefficient and impractical for large-scale or real-time operations~\cite{arul2024and}. Recent frameworks address bandwidth constraints by enabling agents to exchange concise and abstract representations of their states rather than exhaustive sensor data. Methods such as Multi-Agent Reinforcement Learning (MARL) and decentralized robotic perception encode observations (e.g., occupancy maps, critical features, and intent predictions) into compact message vectors for efficient transmission~\cite{arul2024and}. Despite significant communication overhead reductions, these techniques still face challenges in synchronizing heterogeneous multimodal data (e.g., visual, LiDAR, and telemetry) across agents, particularly in dynamic scenarios.

Figure~\ref{fig:semantic_comm_overview} summarizes popular multi-agent frameworks, detailing their supported modalities, typical application scenarios with representative multimodal multi-agent communication scenarios, depicting three standard communication patterns:

\begin{enumerate}
\item \textbf{Intradomain Inter-Agent Communication:} Vehicles of the same type (e.g., self-driving cars) broadcast semantic information such as current traffic state, intended paths, and future motion plans to coordinate actions under shared infrastructure.

\item \textbf{Intradomain Intra-Agent Communication:} Within each autonomous agent, distinct modules (e.g., perception, avoidance planner, and trajectory executor) exchange multimodal summaries internally to enable reactive decision-making. A specific example is the Robot Collaboration presented by Helix AI, where multiple robots with identical vision-language-action (VLA) models internally share semantic representations to collaboratively achieve semantic goals, e.g., “store the bag of cookies in the open drawer.”

\item \textbf{Interdomain Inter-Agent Communication:} Heterogeneous agents from different domains (e.g., delivery vehicles and warehouse robots) share mission status and planning data via structured message formats. These messages often pass through cloud-edge repositories or standardized interfaces to support workflow-level coordination.
\end{enumerate}

The concept of intra-agents is defined as identical models with exactly the same training data and parameters, thus restricting communication strictly within a single domain. As a result, an interdomain intra-agent scenario is not included in the categories.

To support such distributed semantic exchange, frameworks like the Model Context Protocol (MCP) have been proposed. MCP introduces structured ``context blocks'' that standardize inter-agent communication via:

\begin{itemize}
\item \textbf{Metadata Header:} Source ID, timestamps, and priority indicators for content freshness and relevance.
\item \textbf{Semantic Tags:} Categorical descriptors to index and retrieve relevant information.
\item \textbf{Payload Field:} Compact, semantically meaningful content (e.g., summarized states, obstacle maps, or task updates).
\end{itemize}

Agents publish and query these context blocks from shared repositories, using key-based search or dynamic relevance scoring. Synchronization mechanisms such as version control and atomic updates ensure consistency. MCP is designed for interoperability with modern agent architectures including LLMs and hybrid reasoning modules~\cite{qin2024knowledge}.

However, current implementations of MCP and related protocols primarily enforce structure and format, but not adaptive content selection. That is, they do not determine what semantic elements are most relevant to transmit, when to retrieve additional detail, or how to refine shared knowledge dynamically as task conditions evolve.

To address this limitation, we propose a retrieval-augmented semantic communication strategy that supports iterative semantic refinement. The proposed RAMSemCom framework introduces patch-level semantic retrieval and deep reinforcement learning (DRL)-based scheduling to enable bandwidth-aware, adaptive communication across agents. Rather than relying on static message formats, RAMSemCom allows agents to progressively request the most relevant multimodal content as needed, based on real-time task demands and resource constraints. 

\section{Proposed Retrieval-Augmented Semantic Communication Framework}

\subsection{Framework Overview and Design Principles}

RAMSemCom framework is explicitly designed for bandwidth-constrained, multi-agent environments, enabling effective iterative semantic refinement through dynamic multimodal information exchange. Inspired by RAP~\cite{wang2025retrieval}, RAMSemCom systematically combines retrieval mechanisms with semantic communication to optimize multimodal content exchanges among distributed agents. Each component of RAMSemCom plays a critical role within the framework:

\begin{itemize} \item \textbf{Retrieval}: Agents dynamically query additional semantic elements using top-$k$ semantic similarity searches, selectively retrieving essential multimodal patches from semantic databases to significantly reduce transmission overhead. 

\item \textbf{Augmentation}: Retrieved semantic elements are iteratively fused with existing semantic contexts at receiving agents, enriching their knowledge and enhancing decision-making accuracy.  

\item \textbf{Multimodal Semantic Communication}: Meaning-oriented multimodal representations are encoded, transmitted, and decoded effectively, allowing agents to exchange critical information without excessive bandwidth usage. \end{itemize}

To illustrate clearly the framework’s hierarchical structure and operational flow, Fig.~\ref{fig:ramsemcom_architecture} provides a detailed overview.

\begin{figure}[t] 
\centering 
\includegraphics[width=\linewidth]{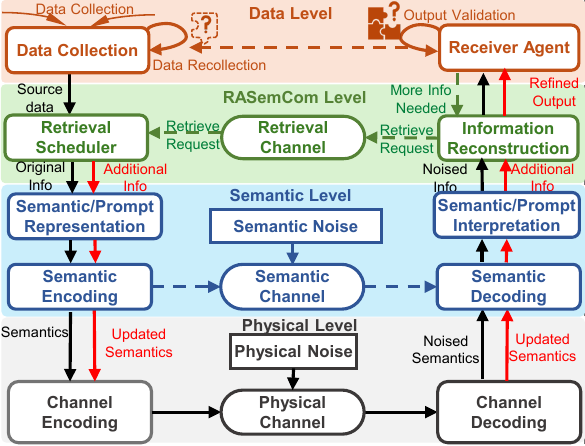} 
\caption{Detailed hierarchical architecture of the RAMSemCom framework, highlighting three distinct layers: Data Level, RAMSemCom Level, Semantic Level, and Physical Level. The figure illustrates the complete process from semantic data encoding, iterative retrieval requests based on semantic uncertainty, and channel-aware retrieval scheduling. Each layer's function is clearly outlined, emphasizing the role of dynamic retrieval scheduling and iterative refinement in minimizing bandwidth usage and ensuring effective semantic interpretation.} 
\label{fig:ramsemcom_architecture} 
\end{figure}

\subsection{Semantic Element Retrieval and Iterative Semantic Refinement}

As shown in Fig.~\ref{fig:retrieval_refinement_process}, RAMSemCom executes an iterative semantic retrieval and refinement process:

\begin{enumerate} \item \textbf{Initial Semantic Transmission}: A sender initially transmits compact semantic summaries of multimodal data. 
\item \textbf{Agent-based Assessment}: Receiving agents assess whether the semantic data sufficiently meets their task-specific requirements. 
\item \textbf{Semantic Retrieval Requests}: When additional information is needed, agents perform semantic similarity searches to identify the most relevant multimodal patches, generating explicit retrieval requests.
\item \textbf{Iterative Refinement}: The retrieved patches are iteratively integrated into the agent’s context, progressively refining semantic understanding until sufficient clarity for task execution is achieved. 
\end{enumerate}

\begin{figure*}[t] \centering \includegraphics[width=0.8\linewidth]{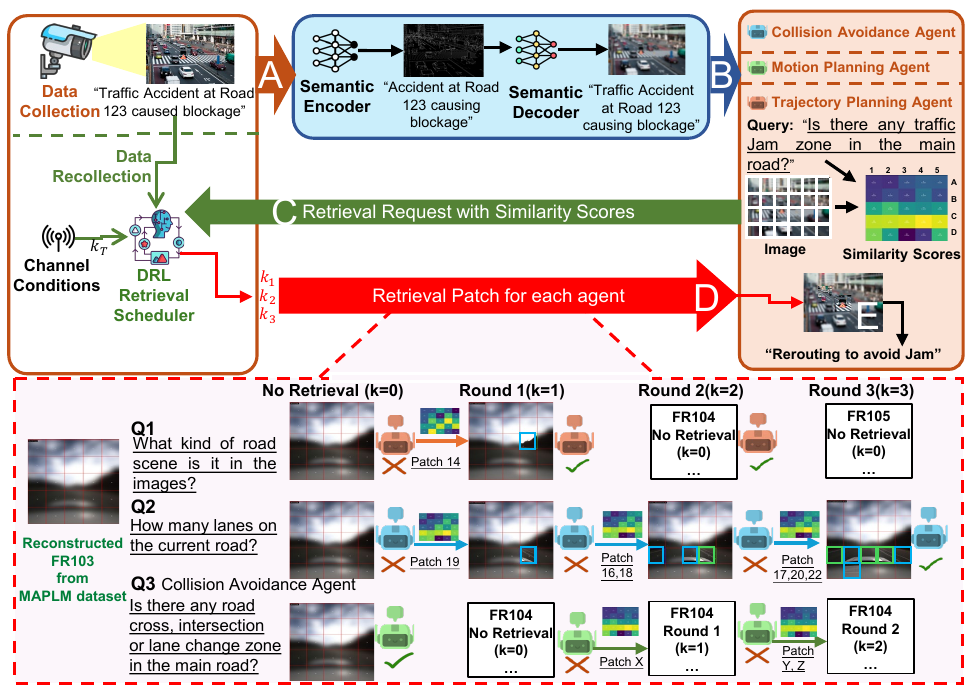} \caption{Visualization of the retrieval-refinement process within RAMSemCom using a representative autonomous driving scenario. Initially, a low-resolution semantic summary of the scene is transmitted (A, B). Agents evaluate task-specific queries (e.g., traffic conditions ahead) and perform top-$k$ semantic similarity searches to identify relevant patches (C). A DRL-based scheduler dynamically allocates retrieval requests considering real-time channel conditions and semantic priorities (D). Iterative retrieval rounds progressively refine agents' semantic perception, leading to accurate decisions such as rerouting to avoid traffic congestion (E).} \label{fig:retrieval_refinement_process} \end{figure*}

\subsection{Dynamic Retrieval Optimization via DRL}

Central to the RAMSemCom’s functionality is the DRL-based optimization strategy for semantic patch retrieval scheduling, which dynamically manages retrieval patch size $k$ based on several criteria:

\begin{itemize} 
\item \textbf{Semantic-Utility Aware Scheduling}: Balances the semantic value of requested patches against associated transmission overhead. 
\item \textbf{Channel-Aware Adaptation}: Adjusts retrieval strategies based on real-time wireless channel conditions, optimizing bandwidth usage. \item \textbf{Latency-Constrained Retrieval}: Ensures timely semantic information retrieval within strict temporal constraints typical in real-time applications. \end{itemize}

The DRL scheduler is strategically positioned within the RAMSemCom Level (as illustrated in Fig.~\ref{fig:ramsemcom_architecture}), providing centralized management of retrieval requests, channel conditions, and bandwidth constraints to ensure effective multi-agent coordination and bandwidth efficiency.

\section{Case Study: Semantic Communication in Multi-Agent Autonomous Driving}

\subsection{Scenario Description: Multi-Agent Bandwidth-Constrained Driving}

Autonomous driving systems critically depend on real-time semantic perception and communication to support safe and efficient navigation. These systems must accurately exchange scene information (e.g., obstacle location, lane conditions, and traffic sign) among intelligent agents (e.g., vehicles and infrastructure units) operating in dynamic environments. However, transmitting high-resolution sensor data (e.g., images and LiDAR point clouds) for each agent is infeasible due to the stringent wireless bandwidth limitations and latency constraints typical in vehicular networks.

To simulate such conditions, we use the MAPLM dataset~\cite{cao2024maplm}, which includes annotated multimodal scenes and standard driving-related Question-Answer (QA) pairs. We randomly select 100 real-world urban driving scenes, each containing relevant visual, LiDAR, and semantic metadata, to form the basis of a retrieval-augmented semantic communication (RAMSemCom) simulation across three autonomous vehicles interacting with a shared roadside infrastructure unit.

Key constraints modeled in this setup include:
\begin{itemize}
    \item \textbf{Shared Bandwidth:} A single 1 MHz wireless channel shared among three vehicles, with an estimated total channel capacity of 5.03 Mbps under 15 dB SNR.
    \item \textbf{Semantic Bottlenecks:} Transmitting full-resolution content for all agents simultaneously exceeds channel capacity, necessitating selective and adaptive communication.
    \item \textbf{Task-Driven Semantics:} Each vehicle must answer a scene-specific QA task based on transmitted content; low semantic fidelity leads to incorrect answers.
    \item \textbf{Retrieval-Based Refinement:} If the initial downsampled image is insufficient, vehicles can request specific high-resolution image patches via top-$k$ retrieval.
\end{itemize}

\subsection{RAMSemCom with DRL-based Optimization}

We integrate a centralized DRL scheduler at the roadside infrastructure to dynamically determine the optimal number of semantic patches ($k$) allocated per vehicle each communication round, adapting intelligently to real-time semantic needs and channel conditions.
Specifically, the DRL scheduler is defined as follows:
\begin{itemize}
    \item \textbf{State}: Includes wireless channel status, semantic similarity rankings of patch requests, and retrieval history.
    \item \textbf{Action}: Determines the discrete number of semantic patches ($k$) allocated individually to each vehicle in the current round.
    \item \textbf{Reward}: Balances successful semantic task completion, bandwidth efficiency, and latency. Higher rewards encourage task accuracy with minimal patch transmission and timely completion.
\end{itemize}

The DRL-driven communication rounds operate iteratively:
\begin{enumerate} 
\item Vehicles initially receive low-resolution semantic summaries and textual context. \item Vehicles locally assess semantic uncertainty, requesting top-$k$ relevant high-resolution patches if needed. 
\item The centralized DRL scheduler allocates optimal patch budgets per vehicle, considering semantic importance and bandwidth availability. 
\item Vehicles iteratively refine their semantic understanding with received patches until task completion or latency limit. \end{enumerate}
This DRL-enhanced RAMSemCom design ensures efficient, timely, and accurate semantic communication among bandwidth-constrained autonomous agents.
\subsection{Numerical Results}
}
\subsubsection{DRL-based Patch Allocation}
\begin{figure}[tpb]
    \centering
    \includegraphics[width=\linewidth]{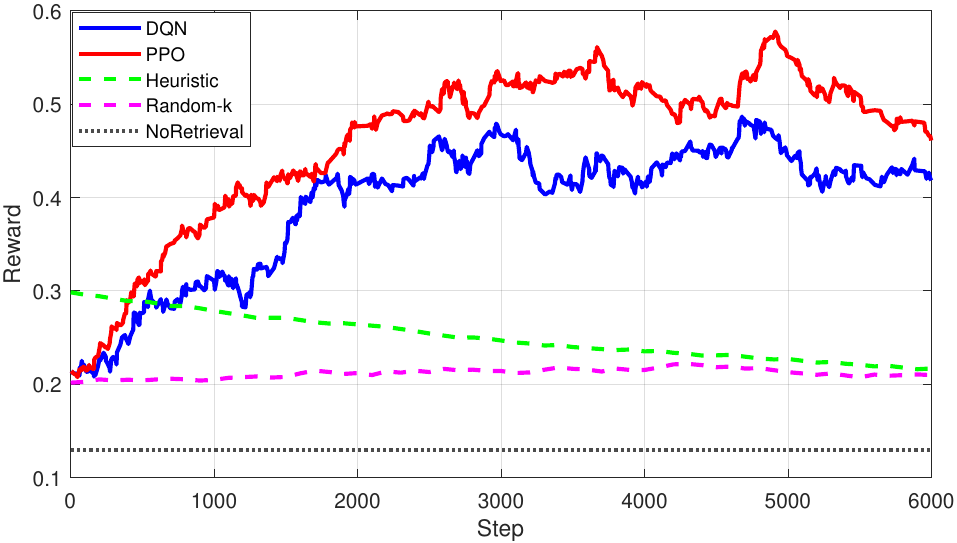}
    \caption{Comparison of reward performance for different patch retrieval strategies over training episodes. The DRL-based methods (PPO and DQN) significantly outperform fixed baselines, demonstrating their effectiveness in dynamically allocating semantic retrieval bandwidth. The NoRetrieval line ($k=0$) serves as a performance floor, while heuristic and random $k$ strategies show limited adaptability in complex task scenarios.}
    \label{fig:reward_curve}
\end{figure}
Figure~\ref{fig:reward_curve} presents the performance comparison of various retrieval scheduling strategies in terms of accumulated semantic rewards. Among all evaluated approaches, the Proximal Policy Optimization (PPO) method consistently achieves the highest reward, reaching a peak of approximately 0.56, followed by Deep Q-Network (DQN), which converges around 0.46. In contrast, the heuristic-based strategy stabilizes below 0.30, while the random $k$ assignment remains near 0.23. The static baseline without any retrieval (i.e., $k=0$) provides a constant reward of 0.13, serving as a lower bound.

These results demonstrate that DRL-based approaches can effectively learn context-aware retrieval policies that optimize task utility under bandwidth constraints. Although the optimization target is the number of semantic patches ($k$), it directly corresponds to channel resource allocation, given that each patch transmission consumes a fixed amount of wireless bandwidth. Therefore, the retrieval scheduling policy can be viewed as a form of discrete channel scheduling across semantic agents.

Compared to DQN, PPO exhibits smoother reward convergence and better long-term performance. This is attributed to its policy-gradient formulation, which allows for more stable updates in the presence of delayed or sparse semantic feedback. In contrast, the baseline methods are unable to adapt to per-round task difficulty and thus fail to allocate retrieval resources effectively.

\subsubsection{Comparison of Task Completion Efficiency}
\begin{figure}[tpb]
    \centering
    \includegraphics[width=\linewidth]{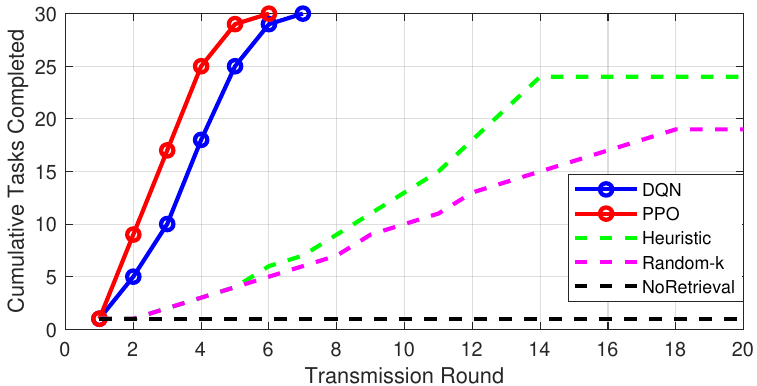}
    \caption{Cumulative number of successfully completed QA tasks as a function of transmission round. DRL-based methods (PPO and DQN) converge significantly faster than heuristic and non-learning baselines, highlighting the effectiveness of adaptive semantic patch allocation under bandwidth constraints.}
    \label{fig:task_curve}
\end{figure}
To assess the efficiency and convergence behavior of different patch allocation strategies, we simulate a multi-agent semantic communication scenario using 30 QA tasks randomly selected from the evaluation dataset. As shown in Figure~\ref{fig:task_curve}, both DRL-based methods PPO and DQN, demonstrate significantly faster completion speed compared to non-learning baselines. PPO achieves full task completion within six communication rounds, while DQN completes all tasks by the seventh round. In contrast, the heuristic and random-k baselines exhibit slower finishing rate, with many tasks remaining incomplete even after fifteen rounds. The No-Retrieval baseline only completes one task due to its lack of semantic refinement. These results highlight the advantages of dynamic retrieval scheduling in improving semantic task efficiency under bandwidth constraints.
\section{Future Directions}
Several promising future research directions can be explored to extend RAMSemCom's effectiveness and applicability:
\begin{itemize}
\item \textbf{Federated Semantic Learning:} Investigate federated approaches to enhance RAMSemCom by enabling distributed semantic learning across agents without sharing raw data, thereby improving model generalizability and privacy preservation.
\item \textbf{Adaptive Semantic Compression:} Develop adaptive compression algorithms integrated with retrieval-based refinement to dynamically balance between data fidelity and bandwidth efficiency based on real-time wireless conditions.
\item \textbf{Cross-Modality Retrieval Optimization:} Explore more sophisticated cross-modal retrieval mechanisms leveraging attention-based transformer architectures to improve the accuracy and efficiency of semantic retrieval across diverse multimodal datasets.
\item \textbf{Scalability and Deployment:} Evaluate the scalability of RAMSemCom in large-scale IoT and smart city applications, addressing practical challenges such as latency management, synchronization across numerous devices, and robust deployment under varying wireless channel conditions.
\end{itemize}

\section{Conclusion}
This paper introduced RAMSemCom, a novel retrieval-augmented multimodal semantic communication framework designed for wireless agentic AI systems operating under bandwidth constraints. Leveraging iterative retrieval-driven semantic refinement and DRL-based dynamic optimization, RAMSemCom significantly reduces transmission overhead while maintaining high semantic fidelity. Our case study in multi-agent autonomous driving scenarios validated the superiority of RAMSemCom over heuristic and non-adaptive approaches, demonstrating faster task convergence and improved resource efficiency. Future research will continue refining RAMSemCom through federated learning approaches, adaptive semantic compression techniques, cross-modality optimization, and scalability evaluations, paving the way for broader real-world deployments.

\bibliographystyle{IEEEtran}
\bibliography{main}

\end{document}